# Epitaxial stabilization and phase instability of $VO_2$ polymorphs

Shinbuhm Lee, Ilia N. Ivanov, Jong K. Keum & Ho Nyung Lee*

Oak Ridge National Laboratory, Oak Ridge, Tennessee 37831, USA
*hnlee@ornl.gov

*This manuscript has been authored by UT-Battelle, LLC under Contract No. DE-AC05-00OR22725 with the U.S. Department of Energy. The United States Government retains and the publisher, by accepting the article for publication, acknowledges that the United States Government retains a non-exclusive, paid-up, irrevocable, world-wide license to publish or reproduce the published form of this manuscript, or allow others to do so, for United States Government purposes. The Department of Energy will provide public access to these results of federally sponsored research in accordance with the DOE Public Access Plan (http://energy.gov/downloads/doe-public-access-plan).*

The VO$_2$ polymorphs, i.e., VO$_2$(A), VO$_2$(B), VO$_2$(M1) and VO$_2$(R), have a wide spectrum of functionalities useful for many potential applications in information and energy technologies. However, synthesis of phase pure materials, especially in thin film forms, has been a challenging task due to the fact that the VO$_2$ polymorphs are closely related to each other in a thermodynamic framework. Here, we report epitaxial stabilization of the VO$_2$ polymorphs to synthesize high quality single crystalline thin films and study the phase stability of those metastable materials. We selectively deposit all the phases on various perovskite substrates with different crystallographic orientations. By investigating the phase instability, phonon modes and transport behaviours, not only do we find distinctively contrasting physical properties of the VO$_2$ polymorphs, but that the polymorphs could be on the verge of phase transitions when heated as low as ~400 °C. Our successful epitaxy of both VO$_2$(A) and VO$_2$(B) phases, which are rarely studied due to the lack of phase pure materials, will open the door to the fundamental studies of VO$_2$ polymorphs for potential applications in advanced electronic and energy devices.

**Introduction**

Vanadium dioxides ($VO_2$) are strongly correlated $d^1$ electron systems and are known to have several polymorphs, which include $VO_2$(A), $VO_2$(B), $VO_2$(M1) and $VO_2$(R). While the chemical formula is the same, their crystalline and electronic structures are completely different and highly complex, exhibiting many interesting electrical, optical and chemical properties owing to the strong correlation[1–3]. Among the aforementioned $VO_2$ polymorphs, the rutile $VO_2$(R) and the monoclinic $VO_2$(M1) have been the most widely studied phases due primarily to their metal-to-insulator transition (MIT) temperature close to room temperature (68 °C)[1–3]. Since this phase transition is accompanied by a huge change in resistivity by three orders of magnitude, $VO_2$(R) and $VO_2$(M1) have attracted tremendous attention for the electronic and optical applications, such as smart windows[4], frequency-agile metamaterials[5,6] and electrical switches[7–9].

The monoclinic $VO_2$(B) phase has also been explored. However, the focus has been on utilization of the open framework, which originates from the edge-sharing $VO_6$ octahedra[10–12]. Such open framework makes $VO_2$(B) a promising energy material, which can be used as electrodes in Li-ion batteries[13]. However, the growth of single crystalline $VO_2$(B) has not been very successful due to the complex crystal structure[14,15]. Similarly, the study of the tetragonal $VO_2$(A) has so far been very limited[16,17] as compared to other $VO_2$ polymorphs, due to the difficulty in synthesizing phase pure crystals. Thus, their physical properties and potential for technical applications have not been much explored.

One of the main reasons for the difficulty in preparing phase pure $VO_2$ polymorphs is the narrow range of phase diagram[3] and, more importantly, the $VO_2$ polymorphs are closely related each another in a thermodynamic framework[10,12,16]. For example, it has been shown that the $VO_2(A)$ and $VO_2(B)$ phases are metastable in bulk and undergo an irreversible phase change into $VO_2(R)$ upon heating[10,12,16], resulting in a mixture of $VO_2$ polymorphs. The formation of such mixed phases hinders the accurate understanding of the physical properties of the $VO_2$ polymorphs. Hence, preparation of phase pure and high quality crystalline materials has been one of the major challenges in $VO_2$ research.

Epitaxial stabilization of crystalline materials by formation of low energy interface is a well-known approach to creating pure phase materials[18–20]. Because the stability of these non-equilibrium materials is affected by both thermodynamic and kinetic factors, the highly non-equilibrium film growth conditions offered by pulsed laser epitaxy (PLE) provides a unique opportunity to discover a wide range of materials with unprecedented functionalities.

Here, we report comparatively the physical properties of four $VO_2$ polymorphs (i.e., R, M1, A and B phases) epitaxially stabilized by PLE on various perovskite substrates with different crystallographic orientations, i.e., $ABO_3(001)$, $ABO_3(011)$ and $ABO_3(111)$. Distinctively contrasting phase stability, lattice motions and transport properties reported here will provide useful information to develop $VO_2$-based electronic devices and energy materials.

**Results and Discussion**

In order to selectively grow $VO_2$ polymorphs, commercially-available perovskite-oxide substrates, including $TbScO_3$ (TSO), $SrTiO_3$ (STO), $(LaAlO_3)_{0.3}(Sr_2AlTaO_6)_{0.7}$ (LSAT), LaAlO$_3$ (LAO) and YAlO$_3$ (YAO), were used. As summarized in Table I, we were able to epitaxially grow (1) the tetragonal VO$_2$(A) phase on (011)-oriented STO and LAO substrates; (2) the monoclinic VO$_2$(B) phase on a wide selection of (001)-oriented substrates, including pseudo-cubic TSO, STO, LSAT, LAO and pseudo-cubic YAO; and (3) the monoclinic VO$_2$(M1) phase on (111)-oriented STO, LSAT and LAO substrates, which commonly have a 3$m$ surface symmetry.

The selective growth occurs due to preferential in-plane lattice matching of perovskite-oxide substrates with the VO$_2$ polymorphs. VO$_2$(B) has a low-symmetry monoclinic structure (space group of C2/m) with lattice constants of $a$ = 12.03 Å, $b$ = 3.69 Å, $c$ = 6.42 Å and $\beta$ = 106.6°, as summarized in Table I and as schematically shown in Fig. 1a. Various X-ray diffraction (XRD) scans, including $\theta$−2$\theta$ scans shown in Fig. 2a and $\phi$ scans (data not shown), for VO$_2$(B) films on (001)STO ($a_{STO}$ = 3.905 Å) confirmed the following epitaxy relationship: (001)VO$_2$(B) || (001)STO; [100]VO$_2$(B) || [100]STO. The lattice mismatch $(a_{sub} - a_{film})/a_{sub} \times 100$ was −2.6 % for [010]VO$_2$(B) || [010]STO and +5.8 % for [100]VO$_2$(B) || [100]STO, where the negative and positive signs indicate compressive and tensile strain, respectively.

VO$_2$(A) has a tetragonal structure (space group of P4$_2$/ncm) with lattice constants of $a$ = $b$ = 8.43 Å and $c$ = 7.68 Å, as schematically shown in Fig. 1b. We found that the single crystalline VO$_2$(A) phase could be grown best on (011)STO with the following epitaxy relationship: (100)VO$_2$(A) || (011)STO; [010]VO$_2$(A) || [011]STO, as confirmed by XRD $\theta$−2$\theta$ scans (see Fig. 2b) and $\phi$ scans (data not shown). The mismatches along the two orthogonal directions, i.e., [010]VO$_2$(A) || [011]STO and [001]VO$_2$(A) || [100]STO are −1.7 and +1.7%, respectively.

The VO$_2$(M1) phase has a low-symmetry monoclinic structure (space group of P2$_1$/c) with lattice constant of $a$ = 5.38 Å, $b$ = 4.52 Å, $c$ = 5.74 Å and $\beta$ = 122.6°, as schematically shown in Fig. 1c. There have been several reports on the successful growth of VO$_2$(M1) films on substrates with a 3$m$ surface symmetry[21] such as (0001)Al$_2$O$_3$, (111)MgAl$_2$O$_4$, (111)MgO and (0001)ZnO. In our study, we mainly attempted to epitaxially grow the films on (111)STO substrates to unify the substrates for VO$_2$ polymorph films. As shown in Fig. 2c, we found that VO$_2$(M1) could be grown on (111)STO with the following epitaxy relationship: (010)VO$_2$(M1) || (111)STO and [010]VO$_2$(M1) || [111]STO. The lattice mismatch is −3.8 % along [001]VO$_2$(M1) || [1$\bar{1}$0]STO and +2.6 % along [100]VO$_2$(M1) || [1$\bar{1}$0]STO.

While the three VO$_2$ phases listed above are accessible at room temperature from as grown films, we also tried to access to the VO$_2$(R) phase via a structural phase transition by heating a VO$_2$(M1) film above the $T_c$ (68 °C). As shown in Fig. 2c, we were able to confirm the phase transition into the VO$_2$(R) phase by performing an XRD $\theta$−2$\theta$ scan at 100 °C, which is higher than the $T_c$. Both VO$_2$(R) and VO$_2$(M1) phases on (111)STO are (010)-oriented.

Among the growth parameters, we found that a proper choice of the substrate temperature, $T_s$, is critical, in particular for VO$_2$(A) and VO$_2$(B) phases on perovskite substrates. As shown in Table I, we could reproducibly grow VO$_2$(A) and VO$_2$(B) phases when $T_s$ was lower than 430 °C. On the other hand, the growth of VO$_2$(M1) phase was quite insensitive to $T_s$ as we confirmed the growth of high quality films in a wide temperature window (400 $\leq T_s \leq$ 600 °C).

To evaluate the thermal stability of VO$_2$ polymorphs, epitaxial films of VO$_2$(A), VO$_2$(B) and VO$_2$(M1) phases were heated up to 600 °C. We kept the samples in vacuum (~0.37 Torr) to avoid spontaneous oxidation into V$_2$O$_5$ phase[22]. Figures 3a and b show the phase evolution of VO$_2$(B)/STO(001) and VO$_2$(A)/STO(011), respectively, characterized by XRD $\theta-2\theta$ scans as a function of temperature. In case of VO$_2$(B) on STO(001), upon heating, XRD peaks corresponding to 00$l$ VO$_2$(B) disappeared above 430 °C and then the 330 VO$_2$(A) peak subsequently appeared above 440 °C, indicating the formation of polycrystalline VO$_2$(A). When we further increased $T_s$, the VO$_2$(A) phase disappeared above 470 °C and the polycrystalline VO$_2$(R) phase appeared above 520 °C. This transformation, i.e., VO$_2$(B) → VO$_2$(A) → VO$_2$(R), indicates that the structural frameworks are similar among the phases. The first transition to A-phase is known to associate with the realignment of VO$_6$ octahedra from edge shared to face shared[10] and, the second transition to the R-phase is attributed to the reorientation of the half of the VO$_6$ octahedra[10].

As shown in Fig. 3b, the VO$_2$(A)/STO(011) also revealed similar thermal stability. The peaks corresponding to $l$00 VO$_2$(A) disappeared above 430 °C and polycrystalline VO$_2$(R) was subsequently formed above 470 °C. The phase transitions of both VO$_2$(B) and VO$_2$(A) were irreversible upon cooling. The thermal instability of VO$_2$(A) and VO$_2$(B) explains the formation of mixed phase VO$_2$ polymorphs with VO$_2$(R) as an impurity phase often observed from films grown above 430 °C. The observation of MIT at 68 °C in VO$_2$(A) and VO$_2$(B) films grown above 430 °C clearly indicates inclusion of VO$_2$(R) as an impurity phase[18]. We note that, on the other hand, the VO$_2$(M1) phase was converted into VO$_2$(R) at ~68 °C upon heating and was stable up to 600 °C (data not shown). Upon cooling, $VO_2(R)$ was converted back to $VO_2(M1)$, indicating a reversible phase evolution with good thermal stability.

Since the $VO_2$ polymorphs have distinct structures, one can expect highly contrasting vibrational characteristics of lattice. Thus, identifying the phonon mode is a good measure of phase purity. In order to comparatively understand the phonon modes, Raman spectroscopy was carried out for the $VO_2$ polymorphs by growing films on LAO substrates. The latter were used because dominant Raman spectral features of LAO are isolated at very low wavelength (32 and 123 $cm^{-1}$)[23]. As shown in Fig. 4, the $VO_2$ polymorphs revealed contrasting Raman spectra compared to each another. As compared to Raman data available from nanostructured materials[24−26], we were able to confirm the phase purity of our epitaxial films.

In addition to the phase confirmation, the Raman spectra from $VO_2$ provide more detailed information about the local structure. There are three sets of V−O modes[27] within wavenumber of 100−1100 $cm^{-1}$. At low wavenumber (< 400 $cm^{-1}$), the bands are assigned to V−O−V bending modes; at intermediate wavenumber (400−800 $cm^{-1}$), the bands are attributed to V−O−V stretching modes; and at high wavenumber (> 800 $cm^{-1}$), the bands are assigned to V=O stretching modes of distorted octahedral and distorted square-pyramids. As shown in Fig. 4a, the phonon modes in epitaxial films of $VO_2(B)$ were mainly observed at low and intermediate wavenumbers (152, 263 and 480 $cm^{-1}$), indicating that bending and stretching modes of V−O−V are dominant in $VO_2(B)$. On the other hand, as shown in Fig. 4b, the phonon modes in $VO_2(A)$ were mainly observed at high and intermediate wavenumbers (152, 485 and 887 $cm^{-1}$), which implies that the stretching modes of V−O−V and V=O are dominant lattice motions in $VO_2(A)$. The phonon modes in VO$_2$(M1) are very complex and composed of stretching and bending of V−O−V and zigzag chains of V−V. The phonon modes in VO$_2$(R) dominantly include stretching modes of V−O−V, which indicates that the crystal structure of VO$_2$(R) is more symmetric than VO$_2$(M1)[26,28,29].

While the transport properties of VO$_2$(M1) and VO$_2$(R) have been extensively studied[1−3,8,9,22,28−30], the physical properties of VO$_2$(B) and VO$_2$(A) phases have not been much explored due to difficulty in preparing phase pure thin films. Figure 5 shows transport characteristics of VO$_2$(B), VO$_2$(A) and VO$_2$(M1) films grown on STO substrates. VO$_2$(A) showed a monotonic decrease of resistivity as increasing the temperature, typical for insulators. While still insulating over the temperature range we measured, VO$_2$(B) revealed more or less semiconducting behaviours with much smaller resistivity compared to that of VO$_2$(A), i.e., $\rho_{VO_2(B)}^{300\,K} \approx 0.02\ \Omega\cdot cm$ and $\rho_{VO_2(A)}^{300\,K} \approx 60\ \Omega\cdot cm$. The resistivity in our VO$_2$(A)/STO(011) is higher than that reported in VO$_2$(A)/STO(001)[20] by one order of magnitude. The reason is unclear, but one can consider that the film on (001)STO is under a different strain state or that the growth on a (001)STO substrate may include a small amount of VO$_2$(B) since their thermal phase boundary is relatively low[10,12,16], as shown in Fig. 3. In the case of VO$_2$(R) phase, we also observed the MIT at 340 K from VO$_2$(M1) to VO$_2$(R) phase change upon heating, similarly observed from many previous studies[1−3,8,9,22,28−30]. The MIT accompanied a sudden decrease of the resistivity by 3−4 orders of magnitude, which is comparable to high quality epitaxial films grown on Al$_2$O$_3$(0001)[22]. This excellent performance could be attributed to the high crystallinity of our epitaxial films ($\Delta\omega < 0.1°$). The transition temperature is consistent with structural phase transition from VO$_2$(M1) to VO$_2$(R), as shown in XRD $\theta-2\theta$ scan in the inset of Fig. 2c. We note that the transport properties of the films grown on LAO substrates were almost identical except for slightly decreased resistivity for films on LAO (data not shown).

Overall, as explained above, the $VO_2$ polymorphs revealed a wide range of electronic ground states, i.e., metal [$VO_2$(R)], semiconductor [$VO_2$(B)] and insulator [$VO_2$(A) and $VO_2$(M1)], depending on their crystal structure. This wide range of electronic ground states makes $VO_2$ highly attractive over other transition metal dioxides, since most other binary oxides have electronic ground states of either metal ($CrO_2$: α-phase and β-phase) or insulator ($TiO_2$: rutile, brookite and anatase). While it is not the main focus of this paper, it is worthy mentioning that Goodenough[28,30] obtained a semiempirical expression for the room temperature critical V−V separation $R_c \approx$ 2.92−2.94 Å for localized and itinerant $3d$ electrons in vanadium oxides.

$R < R_c$ → Itinerant $3d$ electron → Metal;

$R > R_c$ → Localized $3d$ electron → Insulator.

This semiempirical criterion indicates that $VO_2$ polymorphs can be either metal or insulator depending on V−V separation in the distinguishable crystal structures. The $VO_2$(R) phase has a uniform V−V separation of $R$ = 2.87 Å (refs. 28,29), resulting in a metallic ground state. The $VO_2$(M1) phase has zigzag V−V chains of $R$ = 2.60 Å and 3.19 Å (refs. 28,29). The $VO_2$(A) phase has zigzag V−V chains of $R$ = 3.25 Å, 3.11 Å and 2.78 Å (ref. 17). The insulating behaviours that we have observed for those M1 and A-phases are attributed to the localized electrons in one of V−V chains with $R$ = 2.60 Å [$VO_2$(M1)] and 2.78 Å [$VO_2$(A)]. Thus, overall transport behaviours of our epitaxial thin films can be well explained by Goodenough's criterion[28,30]. Since $VO_2$ polymorphs have a wide range of physical properties and, in particular, VO$_2$(B) phase is on the verge of becoming a metal, our report on epitaxial synthesis of high quality thin films can open the door to the discovery of novel phenomena and physical properties by deliberate control of the order parameters by various means, including strain, dimensionality, confinement, etc., which can be accessible via epitaxial heterostructuring.

In conclusion, we grew epitaxial films of VO$_2$ polymorphs. For the growth of phase pure VO$_2$ polymorphs, a careful selection of the growth conditions was necessary especially for the temperature and oxygen pressure. Depending on the crystal orientation of substrates, we found that different phases of VO$_2$ could be selectively grown, i.e., VO$_2$(B)/ABO$_3$(001), VO$_2$(A)/ABO$_3$(011) and VO$_2$(M1)/ABO$_3$(111). Such phases revealed unique phonon modes due to the distinctly different crystal structure and physical properties in spite of the same chemical composition. Since the VO$_2$ polymorphs have a wide range of electronic ground states from metal [VO$_2$(R)] and semiconductor [VO$_2$(B)] to insulator [VO$_2$(A) and VO$_2$(M1)], our epitaxial thin films, which are known to be challenging to grow, will expedite our understanding of underlying physics and developing VO$_2$ polymorphs-based electronic devices utilizing the wide selection of the electronic properties from a single composition.

**Methods**

**Epitaxial film growth.** We deposited epitaxial films of VO$_2$ polymorphs on perovskite oxide substrates by pulsed laser epitaxy. We ablated a sintered VO$_2$ target by a KrF excimer laser (248 nm in wavelength) at a laser fluence of 1 Jcm$^{-2}$ and at a laser repetition rate of 10 Hz. By growing thin films under a wide range of $P(O_2)$ and $T_s$ (10 mTorr < $P(O_2)$ < 25 mTorr and 350 °C < $T_s$ < 600 °C), we found the optimal condition for $VO_2$(A), $VO_2$(B) and $VO_2$(M1), as described in Table 1. It should be noted that $V_2O_3$ is formed at $P(O_2)$ < 10 mTorr and $V_2O_5$ is formed for $P(O_2)$ > 25 mTorr, due to the multivalent nature of vanadium[22].

**Characterization of physical properties.** To investigate the *dc* transport properties, a physical property measurement system (Quantum Design Inc.) was used with Pt contacts in four-probe geometry. X-ray diffraction (XRD) measurements were carried out with a four-circle high-resolution X-ray diffractometer (X'Pert Pro, Panalytical) using the Cu-K$\alpha_1$ radiation and equipped with a hot stage (DHS 900, Anton Paar). High-temperature environmental XRD measurements were conducted under vacuum with base pressure of 0.37 Torr. Raman spectra were recorded at various temperatures using a temperature control stage (LincamScientific Instruments). A Renishaw 1000 confocal Raman microscope was used to measure Raman spectra in back scattering configuration. Each spectrum is a sum average of seven individual spectra taken at different place on the sample through 20× objective. The wavelength of the Raman laser used in these measurements was 532 nm.

**Acknowledgements**

This work was supported by the U.S. Department of Energy, Office of Science, Basic Energy Sciences, Materials Sciences and Engineering Division. The Raman and high temperature XRD measurements were conducted as a user project at the Centre for Nanophase Materials Sciences (CNMS), which are sponsored at Oak Ridge National Laboratory by the Scientific User Facilities Division, U.S. Department of Energy.

**Author contributions**

S.L conceived and designed the experiments under supervision of H.N.L. S.L. fabricated the samples, measured electrical transport and conducted high temperature XRD measurements with help of J.K.K. I.N.I performed Raman spectroscopic measurement. S.L. and H.N.L. wrote the manuscript and other authors reviewed it.

**Additional information**

**Competing financial interests:** The authors declare no competing financial interests.

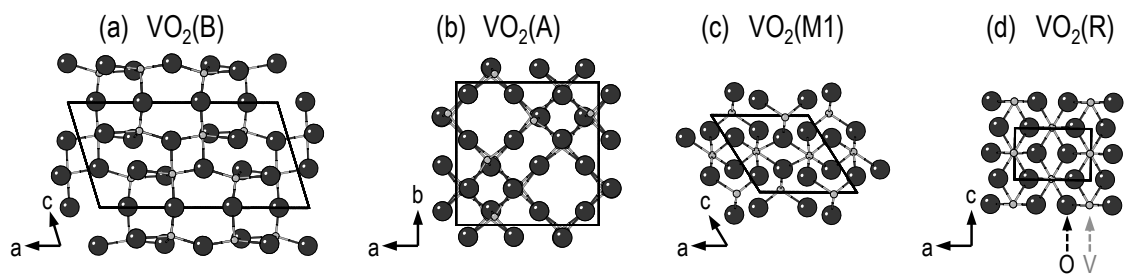

**Figure 1** | Schematics of (**a**) VO$_2$(B), (**b**) VO$_2$(A), (**c**) VO$_2$(M1) and (**d**) VO$_2$(R) phases.

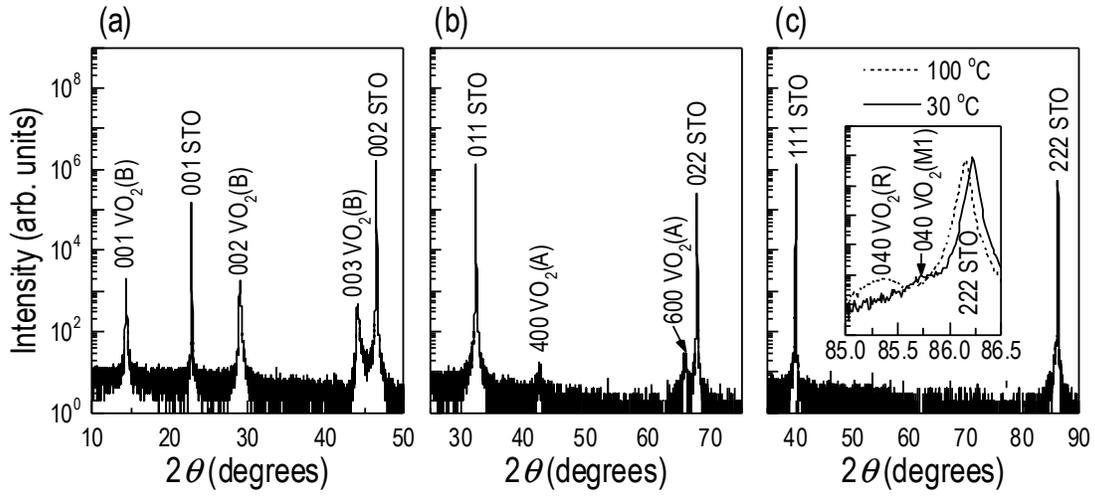

**Figure 2** | XRD $\theta$−$2\theta$ scans of (**a**) $VO_2(B)$, (**b**) $VO_2(A)$ and (**c**) $VO_2(M1)$ thin films on STO (001), (011) and (111) substrates, respectively. The inset in (**c**) shows a XRD scan from the $VO_2(R)$ phase obtained by heating the $VO_2(M1)$ film at 100 $^o$C, which is above the $T_c = 68$ $^o$C.

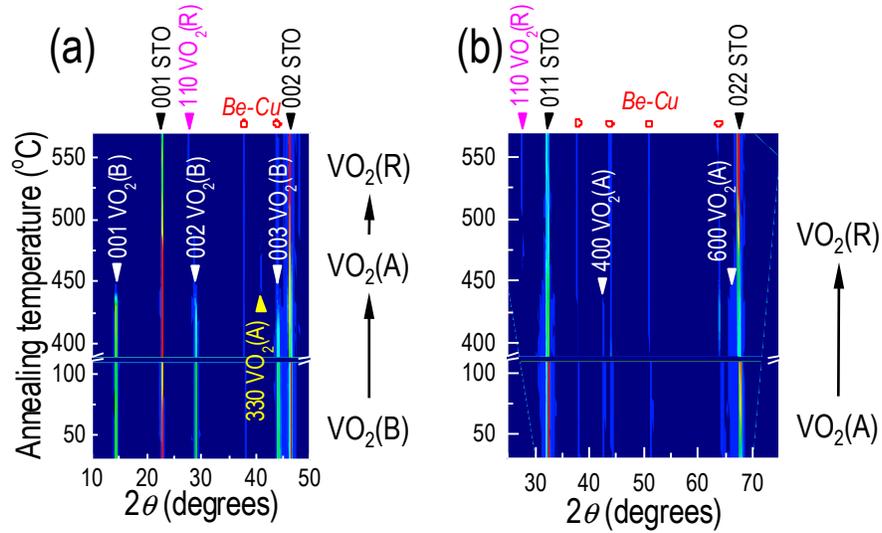

**Figure 3** | Real time XRD $\theta-2\theta$ scans of (**a**) VO$_2$(B)/STO(001) and (**b**) VO$_2$(A)/STO(011) samples as a function of temperature in 0.37 Torr. A clear phase change was observed from both samples, indicating that the phases are in close proximity with each other. The phase changes were, however, irreversible upon cooling.

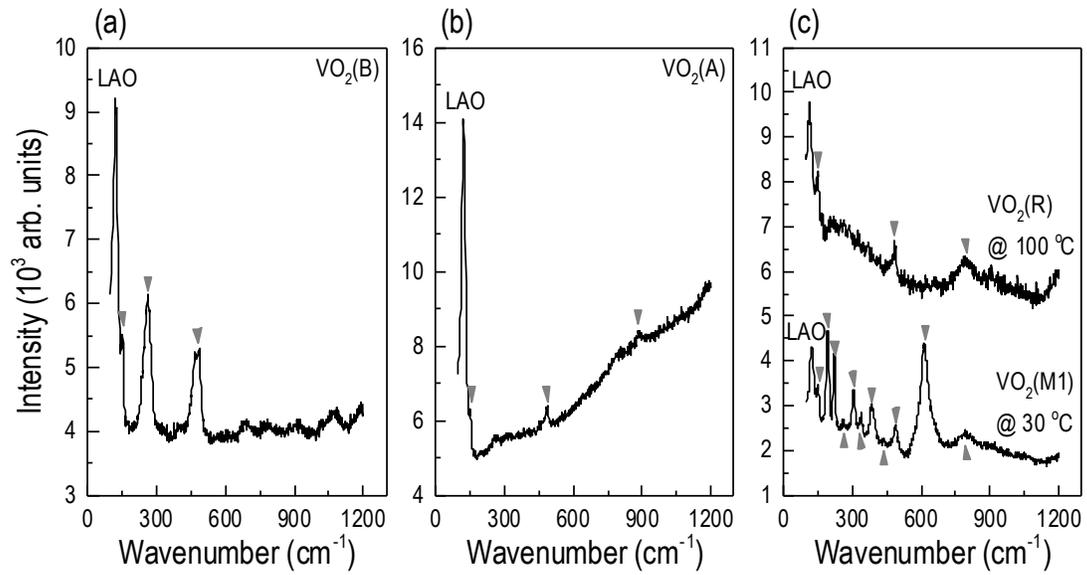

**Figure 4** | Raman spectra of (**a**) $VO_2(B)$, (**b**) $VO_2(A)$, (**c**) $VO_2(M1)$ and $VO_2(R)$ grown on LAO substrates. The spectra were recorded at room temperature except the $VO_2(R)$ phase shown in (**c**), which was obtained by heating the M1 phase sample to 100 $^{\circ}$C in air.

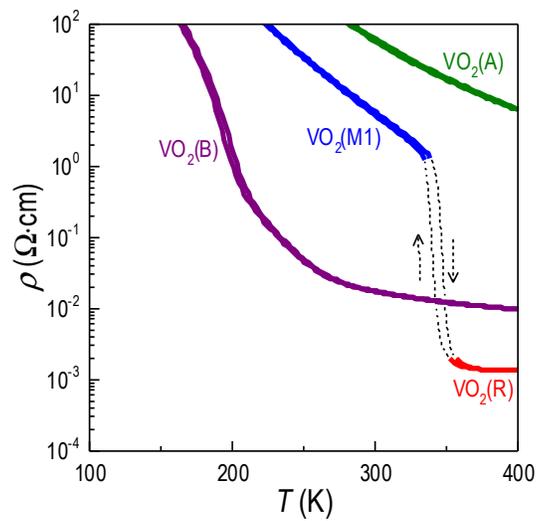

**Figure 5 |** Temperature dependent resistivity for $VO_2(B)$, $VO_2(A)$, $VO_2(M1)$ and $VO_2(R)$ phases grown on STO substrates, exhibiting a distinctly contrasting transport behaviour.

**Table I |** Crystal structure, lattice parameters and growth conditions for $VO_2$ polymorphs.

| $VO_2$ polymorphs | Crystal structure (space group) | Lattice constants in bulk | | | | Substrates for epitaxial growth | Critical growth condition |
|---|---|---|---|---|---|---|---|
| | | a (Å) | b (Å) | c (Å) | β (°) | | |
| $VO_2$(A) | Tetragonal ($P4_2/ncm$) | 8.43 | 8.43 | 7.68 | | STO(011), LAO(011) | $T_s$ < 430 °C |
| $VO_2$(B) | Monoclinic (C2/m) | 12.03 | 3.69 | 6.42 | 106.6 | pc-TSO(001), STO(001), LSAT(001), LAO(001), pc-YAO(001) | $T_s$ < 430 °C |
| $VO_2$(M1) | Monoclinic ($P2_1/c$) | 5.38 | 4.52 | 5.74 | 122.6 | STO(111), LSAT(111), LAO(111) | Not critical to $T_s$ |
| $VO_2$(R) | Tetragonal ($P4_2/mmm$) | 4.55 | 4.55 | 2.88 | | Thermal heating of $VO_2$(M1) above 68 °C | |